\pdfoutput=1
\documentclass{INTERSPEECH2023}

\interspeechcameraready 

\usepackage{glossaries}
\usepackage{subcaption}
\usepackage{multirow}
\usepackage{pifont}
\usepackage{tikz}
\usepackage{pgfplots}
\usepackage{csquotes}
\usepackage{balance}
\usepackage{cite}

\usetikzlibrary{calc}
\usetikzlibrary{tikzmark}
\usetikzlibrary{arrows, positioning}
\usetikzlibrary{fit}
\usetikzlibrary{patterns}
\usepgfplotslibrary[groupplots]
\pgfplotsset{compat=newest}
\usepackage{cleveref}

\newacronym{AAM}{AAM}{Additive Angular Margin}
\newacronym{EER}{EER}{Equal Error Rate}
\newacronym{PIT}{PIT}{Permutation Invariant Training}
\newacronym{TAP}{TAP}{Time Average Pooling}
\newacronym{MSE}{MSE}{Mean Squared Error}
\newacronym{DCF}{DCF}{Detection Cost Function}
\newacronym{ROC}{ROC}{Receiver Operating Characteristic}

\usepackage{tabularx, etoolbox, booktabs}
\usepackage{multirow} 

\makeatletter
\renewcommand\section{\@startsection {section}{1}{\z@}%
                                   {-3.25ex \@plus -1ex \@minus -1.2ex}%
                                   {2.05ex \@plus.2ex}%
                                   {\normalfont\Large\bfseries}}
\renewcommand\subsection{\@startsection{subsection}{2}{\z@}%
                                     {-2.ex\@plus -1ex \@minus -.2ex}%
                                     {1.ex \@plus .2ex}%
                                     {\normalfont\large\bfseries}}
\makeatother

\title{A Teacher-Student approach for extracting informative speaker embeddings from speech mixtures}
\name{Tobias Cord-Landwehr$^1$, Christoph Boeddeker$^1$, C\u{a}t\u{a}lin Zoril\u{a}$^2$, Rama Doddipatla$^2$, Reinhold Haeb-Umbach$^1$}
\address{
  $^1$Paderborn University, Department of Communications Engineering, Paderborn, Germany\\
  $^2$Toshiba Cambridge Research Laboratories, Cambridge, United Kingdom }
\email{cord@nt.upb.de, boeddeker@nt.upb.de, catalin.zorila@toshiba.eu, rama.doddipatla@toshiba.eu, haeb@nt.upb.de}

\begin{document}

\maketitle
 
\begin{abstract}
We introduce a monaural neural speaker embeddings extractor that computes an embedding for each speaker present in a speech mixture. To allow for supervised training, a  teacher-student approach is employed: the teacher computes the target embeddings from each speaker's  utterance before the utterances are added to form the mixture, and the student embedding extractor is then tasked to reproduce those embeddings from the speech mixture at its input. 
The system much more reliably verifies the presence or absence of a given speaker in a mixture than a conventional speaker embedding extractor, and even exhibits comparable performance to a multi-channel approach that exploits spatial information for embedding extraction. 
Further, it is shown that a speaker embedding computed from a mixture can be used to check for the presence of that speaker in another mixture. 
\end{abstract}
\noindent\textbf{Index Terms}: speaker embeddings, teacher-student, multi-speaker, speaker identification, monaural

\section{Introduction}
Speaker embeddings are meant to represent the characteristics of a speaker, while being insensitive to what has been spoken. The embeddings are computed from segments of speech in such a way that they exhibit low intra-speaker and large inter-speaker distance.  The current state of the art are neural speaker embedding extractors. They map the input speech segment to the latent space of embedding vectors by means of a neural network, which can be trained using a contrastive  \cite{18_Wan_contrastive_dvectors} or classification-based loss \cite{19_liu_aam_sv,20_Desplanques_ecapa_tdnn,21_Zhou_resnet}.
Since the number of speakers seen during training is much larger than the dimension of the latent space, the model is forced to encode the speaker characteristics rather than memorizing the speaker labels in the speaker embeddings. In this way, speaker embedding  extractors generalize well to speakers unseen during training \cite{18_snyder_xvector}. 
Speaker embedding extractors are commonly employed either as auxiliary systems for speech enhancement \cite{21_zmolikova_extraction} or automatic speech recognition \cite{Saon13ivectorASR}, as part of a diarization system \cite{21_Raj_libricss_system}, or as standalone systems for the (re-)identification and recognition of speakers \cite{22_Brown_VoxSRC}.

An assumption underlying most current systems is that each processed audio segment contains only a single speaker of interest.
If this assumption is not fulfilled, it is well known that the quality of such speaker embeddings degrades, and that the resulting speaker embedding is at most able to represent one of the speakers in the mixture.
Because of this, diarization systems that employ speaker embedding extractors typically discard regions that contain speech overlap \cite{bredin21_interspeech_resegmentation}, or they use an additional uncertainty state to account for the lower reliability of speaker information extracted from those regions of speech \cite{21_Raj_ovsc,20_Landini_VBx}. Then, the speakers active in those regions often are inferred from the context.
The extraction of reliable speaker embeddings thus rests on the availability of regions where the speaker of interest is active alone. Even state of the art diarization systems like the TS-VAD \cite{20_Medennikov_tsvad} have this dependency on long single-speaker regions for proper initialization  \cite{21_Wang_voxsrc_dia}. 

In highly dynamic situations, e.g. in 
informal meetings or in situations where multiple separate conversations are held in parallel, the requirement that each speaker is at least once solely active, cannot be fulfilled easily.
Here, at least some speakers will not 
appear
alone, so that a 
typical
speaker embedding extractor cannot be used to extract a representation of those speakers.
One way to approach this issue is to employ a source separation module as a preprocessor to the standard
speaker embedding extractor~\cite{20_microsoft_voxsrc}, which, however, can introduce additional artefacts into the signal. Alternatively, a multi-channel system is proposed in \cite{21_He_multispeaker_embeddings}, which computes features for each 
sound direction of arrival to detect and identify speakers. 

In this work, we neither require a source separation front-end nor multi-channel input. Instead, we aim to directly compute from a mixture of two speech signals their respective speaker embeddings.
In a sense, this is similar to source separation since we extract two embeddings from the input mixture. However, there is a clear difference: unlike 
the 
speech signals, the speaker embeddings 
cannot be assumed to linearly superpose
in a mixture. 
As a consequence, a specific setup is required for training.
We employ the teacher-student approach introduced in our previous work~\cite{23_cordlandwehr_ts_embeddings_framewise}:
First, a conventional speaker embedding extractor is trained on single-speaker utterances. Then, this model is employed as a teacher for the training of a student embedding extractor. 
The student's objective is to predict those embedding vectors from a segment of speech mixture at its input.
One way to view this setup is that the teacher 
defines the latent space of speaker embeddings, while the student inherits this space and is only tasked to separate a  mixture of two speakers into their respective representations in that space.
Contrary to the Wavesplit source separation model \cite{21_Zeghidour_Wavesplit}, where a speaker representation is jointly learned with the separation task, the latent space of the speaker embeddings already is explicitly provided by the teacher.
Therefore, similar to the problem of source separation, we view this system as a speaker embedding separation approach. 

This work is structured as follows:
We start with a description of both the teacher and student models in \cref{sec:speaker_embeddings}. 
Then, \cref{sec:speaker_verification} describes the task of multi-speaker verification and how we trace it back to the classical speaker verification task.
In \cref{sec:evaluation}, the proposed model is then evaluated against a 
typical
single-speaker embedding extractor on a multi-speaker verification task.
Finally, we compare our 
system against a multi-speaker embedding extractor on realistic meeting recordings. 

\section{Teacher-student based  embedding extraction}
\label{sec:speaker_embeddings}
The proposed system for multi-speaker embedding extraction consists of a single-speaker teacher that is used to provide training targets for a multi-speaker student model.
First, the teacher model is trained with single-speaker utterances. Then, the teacher model is fixed, and  the student network is trained with speech mixtures at its input and  the speaker embeddings of the teacher as  training targets.

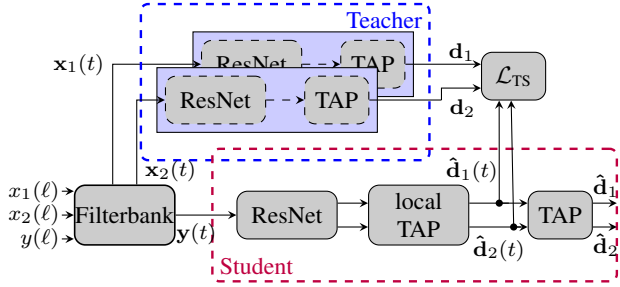
\begin{figure}[bt]
\centering
%
\begin{tikzpicture}[semithick,auto,
block_high/.style={
		rectangle,
		draw,
		fill=black!20,
		text centered,
		text width=3.5em,
		rounded corners,
 		minimum height=2.5em,
		minimum width=3.5em},
block/.style={
	rectangle,
	draw,
	fill=black!20,
	text centered,
	text width=3.5em,
	rounded corners,
	minimum height=2em,
	minimum width=3.5em},
 block_small/.style={
	rectangle,
	draw,
	fill=black!20,
	text centered,
	text width=2em,
	rounded corners,
	minimum height=2em,
	minimum width=2em},
block_teacher/.style={
		rectangle,
		draw,
		dashed,
		fill=black!20,
		text centered,
		text width=3.5em,
		rounded corners,
 		minimum height=2em,
		minimum width=3.5em},
block_teacher_small/.style={
		rectangle,
		draw,
		dashed,
		fill=black!20,
		text centered,
		text width=2em,
		rounded corners,
 		minimum height=2em,
		minimum width=2em},
mul/.style={
        circle,
        draw,
    },		
	]
\tikzset{>=stealth}
\tikzstyle{branch}=[{circle,inner sep=0pt,minimum size=0.3em,fill=black}]

\tikzset{pics/.cd,
	pic switch closer/.style args={#1 times #2}{code={
		\tikzset{x=#1/2,y=#2/2}
		\coordinate (-in) at (1,0);
		\coordinate (-out) at (-1,0);
		
		\draw [line cap=rect] (-1, 0) -- ++(0.1,0) -- ++(20:1.9);
		
	}}
}

    \pgfdeclarelayer{background1}
    \pgfdeclarelayer{background2}
    \pgfdeclarelayer{mid1}
    \pgfdeclarelayer{mid2}
    \pgfdeclarelayer{foreground}
    \pgfsetlayers{background1,background2,mid1,mid2,main,foreground}

    \node[block_high] (encoder) {};
    \node (encoderlabel) at ($(encoder)$) {Filterbank};

	\begin{pgfonlayer}{background2}
		\node [block_teacher] (teacherbg) at ($(encoder.east) + (1cm, 2cm)$) {ResNet};
		\node [block_teacher_small, right=0.5cm of teacherbg] (poolingbg) {TAP};
	\end{pgfonlayer}
	\begin{pgfonlayer}{background1}
		\node [draw=black, fill=blue!20, fit={(teacherbg) (poolingbg)}] {};
	\end{pgfonlayer}

	\begin{pgfonlayer}{mid2}
		\node [block_teacher, anchor=north west] (teacherfg) at ($(teacherbg.north west) + (-1.5em, -1.5em)$) {ResNet};
		\node [block_teacher_small, right=0.5cm of teacherfg] (poolingfg) {TAP};
	\end{pgfonlayer}
	\begin{pgfonlayer}{mid1}
		\node [draw=black, fill=blue!20, fit={(teacherfg) (poolingfg)}] {};
	\end{pgfonlayer}
	    \tikzstyle{box} = [draw, dashed, inner xsep=1em, inner ysep=1.5em, line width=0.1em, rounded corners=0.3em]
		\node [box, draw=blue,fit={(teacherbg) (poolingbg) (teacherfg) (poolingfg)}, label={[anchor=north east, align=left]north east:\color{blue}Teacher}] {};
    \node [block, right=0.8cm of encoder] (student) {ResNet};
    \node [coordinate, right=2cm of student] (splitframe) {};
    \node [branch] (splitframe1) at ($(splitframe) + (0.4em, 0.5em)$) {};
    \node [branch] (splitframe2) at ($(splitframe) + (1.0em, -0.5em)$) {};
    \node [block_small, right=0.5cm of splitframe] (poolingstudent) {TAP};
    \node [block_small] (loss) at ($(poolingbg.east) + (4.5em, -0.5em)$) {$\mathcal{L_{\text{TS}}}$};
    \node[coordinate, right=0.5cm of encoder] (splitteacher) {};
    \node[block, right=0.4cm of student] (shortpooling) {local TAP};
	\node [box, draw=purple,fit={(shortpooling) (student) (poolingstudent)}, label={[anchor=south west, align=left]south west:\color{purple}Student}] {};
   \draw[<-] ($(encoder.west) + (0,1em)$) node[xshift=-0.25em,left] {\footnotesize{$x_1(\ell)$}} -- +(-0.5em, 0);
   \draw[<-] ($(encoder.west)$) node[xshift=-0.25em,left] {\footnotesize{$x_2(\ell)$}} -- +(-0.5em, 0);
   \draw[<-] ($(encoder.west) + (0,-1em)$) node[xshift=-0.25em,left] {\footnotesize{$y(\ell)$}} -- +(-0.5em, 0);
   \begin{pgfonlayer}{background1}
   	\draw[->] ($(encoder.north) + (-0.5em, 0)$) |- node[midway,left] {\footnotesize{$\mathbf{x}_1(t)$}}  ($(teacherbg.west) + (-0em,0)$);
   \end{pgfonlayer}
   \draw[->] ($(encoder.north)+ (0.5em, 0)$)  |- node[near start, right, yshift=-1.2em]  {\footnotesize{$\mathbf{x}_2(t)$}} ($(teacherfg.west) + (-0em,0)$);
   \draw[->] ($(encoder.east)$) --node[below, xshift=-0.4em] {\footnotesize{$\mathbf{y}(t)$}} (student);
   \draw[->] ($(poolingbg.east) +(0.35em,0)$) -- node[at end, above, xshift=-0.8em,yshift=-0.2em] {\footnotesize{$\mathbf{d}_1$}}($(loss.west) +(0,0.5em)$);
   \draw[-] ($(poolingfg.east) +(0.35em,0)$) -- node[near end,below right, xshift=0.65em, yshift=0.4em] {\footnotesize{$\mathbf{d}_2$}} ($(poolingfg.east) +(3em,0)$);
   \draw[->, dashed] (teacherbg) -- (poolingbg);
  \draw[->, dashed] (teacherfg) -- (poolingfg);
  \draw[-, dashed] (poolingbg) -- ($(poolingbg.east) +(0.35em,0)$);
 \draw[-, dashed] (poolingfg) -- ($(poolingfg.east) +(0.35em,0)$);
   \draw[->] ($(poolingfg.east) +(3em,0)$) |-  ($(loss.west)+(0,-0.5em)$);
   \draw[->] ($(student.east)+(0,-0.5em)$) -- ($(shortpooling.west)+(0,-0.5em)$);
   \draw[->] ($(student.east)+(0,0.5em)$) -- ($(shortpooling.west)+(0,0.5em)$);

   \draw[-] ($(shortpooling.east)+(0,-0.5em)$) -- node[near end, above, xshift=-1.5em, yshift=1.5em] {\footnotesize{$\mathbf{\hat{d}}_1(t)$}} ($(splitframe)+(1.3em,-0.5em)$);
   \draw[-] ($(shortpooling.east)+(0,0.5em)$) --  node[near end, below, yshift=-1em] {\footnotesize{$\mathbf{\hat{d}}_2(t)$}} ($(splitframe)+(0.8em,0.5em)$);
   \draw[->] ($(splitframe1)$) -- ($(loss.south) +(-0.6em,0)$);
   \draw[->] ($(splitframe2)$) -- ($(loss.south) +(-0.1em,0)$);
   \draw[->] ($(splitframe1)$) -- ($(poolingstudent.west) +(0,0.5em)$);
   \draw[->] ($(splitframe2)$) -- ($(poolingstudent.west) +(0,-0.5em)$);
   \draw[->] ($(poolingstudent.east) +(0,0.5em)$) --  node[near end, above, xshift=-0.2em, yshift=-0.1em] {\footnotesize{$ \mathbf{\hat d}_1$}}($(poolingstudent.east) +(1em,0.5em)$);
   \draw[->] ($(poolingstudent.east) +(0,-0.5em)$) -- node[near end, below, xshift=-0.2em, yshift=0.1em] {\footnotesize{$ \mathbf{\hat d}_2$}} ($(poolingstudent.east) +(1em,-0.5em)$);

\end{tikzpicture}%
%
\caption{Block diagram of the proposed teacher-student training for multi-speaker embedding extraction}%
\label{fig:teacher_student_model}%
\end{figure}

\subsection{Single-speaker embedding extractor}
Essentially, any speaker embedding extractor can be used as a teacher. In this work, a ResNet34-based d-vector system \cite{21_Zhou_resnet} is employed, which is a widely used architecture for speaker embedding extractors \cite{22_Chen_sjtu_voxsrc,22_Qin_simam_resnet,22_Thienpodt_score_shift_resnet}. 
First, \num{80}-dimensional log mel filterbank features $\mathbf{x}(t)$, $t \in \{1,\ldots , T\}$, with frame index $t$ and frames per utterance $T$ are extracted from an utterance $x(\ell)$
with time domain sample index  $\ell$.
These features are  passed through the ResNet to obtain frame-wise embeddings $ \mathbf d(t)$. Then, the embeddings 
 are aggregated by a \gls{TAP} to obtain a single, $E$-dimensional d-vector $\mathbf d$ for the utterance.
During training, an additional fully connected layer is used to predict the speaker label $c$ from this d-vector.
An \gls{AAM}-Softmax loss \cite{18_wang_aam} 
\begin{align}
\cos{\Theta_c} &= \frac{\mathbf{d}^\mathrm{T}\mathbf{w}_c}{ |\mathbf{d}||\mathbf{w}_c|} \\[-0.2em]
\mathcal{L}_{\mathrm{AAM}} &= 
-\log\frac{e^{s(\cos{\Theta_c}-a)}}{e^{s(\cos{\Theta_c}-a)} + \sum_{c'=1, c' \neq c}^{C} e^{s\cos{\Theta_{c'}}}}
\end{align}
is used as a training criterion. 
Here, $\Theta_c$ denotes the angle between the d-vector $\mathbf d$ and the prototype embedding of the corresponding speaker which is defined by the $c$-th column $\mathbf{w}_c$ of the weight matrix of the fully connected layer. $C$ represents the total number of speakers during training.
This loss is a softmax cross-entropy loss, where the hyperparameters $s>1$ and $a>0$ modify the loss contributions such that the model 
builds more compact clusters after convergence, 
which improves the overall system performance \cite{19_liu_aam_sv}. 
After training, this single-speaker model is kept fixed and the d-vectors $\mathbf d$ are used as targets to train the student.

\subsection{Multi-speaker embedding extraction}
For the training of the multi-speaker student system, a speech mixture $y(\ell)$ 
consisting of $K$ single speaker source signals $x_k(\ell)$ (here, $K=2$) is used as input for the system. Then, the student is tasked to reproduce the d-vectors that are extracted by the teacher from the single speaker utterances $x_k(\ell)$. 

The multi-speaker student's architecture is in large parts a mirror of the teacher. First, $80$-dimensional log mel filterbank features $\mathbf{y}(t)$ are extracted from the speech mixture.
Then, they are passed through a ResNet34 with a $K$ times larger output dimension
and are rearranged
to obtain frame-wise speaker embeddings $\mathbf{\hat{d}}_k(t)$ for each speaker.
Additionally, these embeddings are then averaged by a local \gls{TAP} layer with a size of \num{11} and a stride of \num{1} to encode context information into each embedding. 
Then, a \gls{MSE}-based similarity loss
\begin{align}
\mathcal{L}_{\mathrm{TS}} = \frac{1}{KT}\smash{\sum_t}\min_{\mathbf{\pi}\in\mathcal P}\smash{\sum_k} \lVert \mathbf{d}_k - \mathbf{\hat{d}}_{\pi_k}(t) \rVert^2
\label{eq:l_ts}
\end{align}
is computed between each frame-wise student embedding $\mathbf{\hat d}_k(t)$ and the teacher embeddings $\mathbf{d}_k$ obtained from the single-speaker observations. 

To account for the arbitrary output order of the student embeddings, a \gls{PIT} loss \cite{16_Yu_tPit, 17_Kolbaek_uPit} is used, which assigns the best permutation $\mathbf{\pi}$  between target and estimated d-vectors for each frame (tPIT).
Alternatively, the permutation can be kept constant for the complete utterance by moving the $\min$ operation in \eqref{eq:l_ts} in front of the sum over the frames $t$, which is known as uPIT.
During inference, an additional \gls{TAP} layer aggregates the frame-wise student embeddings into a single embedding $\mathbf{\hat d}_k$ per speaker.

With this teacher-student training, instead of having to learn a latent space as well as to separate both active speakers from each other, the student is trained to directly reproduce the single speaker embeddings of the teacher, because the latent speaker embedding space  is already defined by the teacher.
This simplifies the problem from one of learning a descriptive speaker space while separating speakers from each other to a task of speaker embedding separation into an already known, latent space.
An illustration of the complete teacher-student model is depicted in \cref{fig:teacher_student_model}.

\section{Speaker verification for overlapping speech}
\label{sec:speaker_verification}
Typically, the task of speaker verification \cite{21_nist_sre_plan} consists of computing the similarity of two single-speaker utterances and deciding whether they belong to the same speaker.
After obtaining these similarity scores for a complete trial set, they are aggregated into a single list and compared against the target labels. Then,  the \gls{EER} and \gls{DCF} are calculated based on the scores and labels 
to evaluate the speaker verification performance \cite{21_nist_sre_plan}.

However, this task is no longer clearly defined if one or both of the observations in such a trial pair contain more than one speaker.
Similar to \cite{21_He_multispeaker_embeddings} 
we extend the classical verification task \enquote{single speaker vs. single speaker} (\textit{s~vs.~s}) by two additional scenarios to also consider speech mixtures: 
\begin{itemize}[noitemsep,topsep=0pt]
\item single speaker vs. mixture (\textit{s~vs.~m})
\item mixture vs. mixture (\textit{m~vs.~m}).
\end{itemize}
For  multi-speaker embedding extraction, the \textit{s~vs.~m} setting is the most relevant scenario.
In most settings, e.g. a meeting or conversation, each speaker is  the sole active speaker at least at some point in time.
Therefore, it can be assumed that these
single-speaker 
regions can be used to extract embeddings that are then taken as reference to verify whether this speaker is active in regions containing overlapping speech  \cite{22_Aloradi_embedding_verification}.
On the other hand, the \textit{m~vs.~m} scenario is relevant for scenarios, where even this constraint cannot be fulfilled, which may happen, e.g., if multiple simultaneous discussions are ongoing at the same time. 

\begin{figure}[bt]
\centering
\input{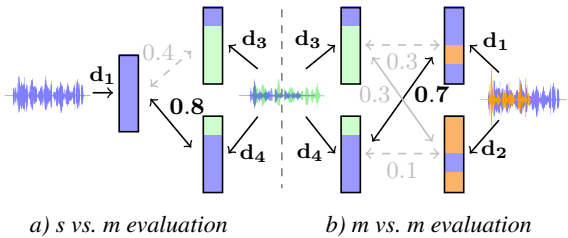}%
\caption{Multi-speaker verification process. For each active speech component, an embedding is extracted, and all pairwise similarities are computed. Only the maximal score is retained for \enquote{any spk} evaluations. For the \enquote{per spk} case, the remaining embedding pair also is evaluated in the m~vs~m scenario.
}
\label{fig:multispeakertrials}
\end{figure}

First, we evaluate both scenarios with the goal to determine whether \textit{any speakers} are matching between both examples in a trial pair. 
Here, from each observation embeddings are extracted as depicted in \cref{fig:multispeakertrials}, and all pairwise similarity scores between the first half and the second half of a trial pair are calculated, i.e., \num{2} and \num{4} in the \textit{s~vs.~m} and the \textit{m~vs.~m} setting, respectively, for $K=2$.
Then, only the maximal similarity score is retained for evaluation.
The calculation of \gls{EER} and \gls{DCF} then is done exactly as for the single-speaker case.
Since only a single similarity score is evaluated, typical single-speaker embedding extractors can also be evaluated in this setup, even though they are not designed for it.

For the \textit{m~vs.~m} scenario, in addition to this \enquote{\textit{any spk}} evaluation, the mixtures are compared \textit{per speaker} to measure how well both speakers of a mixture can be represented. 
Here, the number of positive target labels is determined through the number of identical speakers between both mixtures. These labels are then assigned to the embedding pairs
in a greedy fashion beginning by the highest similarity and excluding assigned pairs for following labels.
Therefore, for each trial pair, $K$ scores and target labels are obtained, as opposed to a single one in the \enquote{\textit{any spk}} setting.
They are then again aggregated and evaluated as in the \textit{s~vs.~s} evaluation to compute both the \gls{EER} and \gls{DCF}.

\section{Evaluation}
\label{sec:evaluation}
\subsection{Training setup}
For training of both the teacher and the student model, \SI{4}{\second} long segments of the VoxCeleb corpus  \cite{19_Nagrani_voxceleb} are used. For filterbank feature extraction, a window size of \SI{20}{\milli\second} and a frame advance of \SI{8}{\milli\second} are chosen.
The teacher is trained on single-speaker utterances augmented with noise from the MUSAN corpus \cite{15_Snyder_MUSAN} and room impulse responses simulated according to \cite{79_allen_image}. 
The student is trained on speech mixtures simulated with MMS-MSG \cite{cordlandwehr2022mms_msg} that consist of two speakers mixed with a power ratio between \SIrange{-5}{5}{\decibel}.
During training, these utterances are cut to the \textit{min} scenario, i.e. the longer utterance in the mixture is cut to match the length of the shorter.
Again, the same data augmentation as for the teacher is used.
Additionally, after \num{10} epochs of training, the target embeddings provided by the teacher are replaced with  embeddings computed from different utterances of the same speaker to make the model more robust against remaining  speaker-unrelated content. 

\subsection{Evaluation sets}
For evaluation, the VoxCeleb1-O trial set \cite{19_Nagrani_voxceleb} is extended by additional examples so that the \textit{s~vs.~m} and\textit{ m~vs.~m} scenarios  can be evaluated\footnote{Trials are available at https://zenodo.org/record/7683872}.
Here, still at most one speaker is identical in the two halves of a trial pair
as to keep them as close as possible to the original VoxCeleb1-O trial set 
and not introduce the number of identical speakers as an additional design parameter in the trial sampling.

Additionally, the proposed model is evaluated on the SSLR database \cite{18_He_SSLR}. This database was originally designed for source localization and consists of re-recordings of AMI meetings \cite{05_Kraaij_AMI}. It was already used in \cite{21_He_multispeaker_embeddings} for the evaluation of multi-channel multi-speaker embeddings with the same scenarios as described in \cref{sec:speaker_verification}.
All models are evaluated w.r.t. \gls{EER} and \gls{DCF}. For \gls{DCF} calculation, a prior probability of \num{0.01} \cite{21_nist_sre_plan} is chosen for the single speaker evaluation, and of \num{0.05} for the multi-speaker verification. 
\begin{table}[bt]
    \centering
        \caption{Speaker verification performance of the teacher model on the VoxCeleb1-O trial set (s~vs.~s)}
    \label{tab:teacher}
    \setlength{\tabcolsep}{4pt}
    \begin{tabular}{l c c c }
        \toprule
        Model &  $E$ &EER [\si{\percent}] & DCF\\
        \midrule
        ECAPA-TDNN \cite{20_Desplanques_ecapa_tdnn} & 1024  & 0.87 & 0.11\\ 
        ECAPA-TDNN \cite{20_Desplanques_ecapa_tdnn} & 512  & 1.01 & 0.13\\ 
        \midrule
        ResNet34 & 256 & 1.06 & 0.16\\
        \bottomrule
    \end{tabular}
\end{table}
\subsection{Performance of the teacher model}
The proposed teacher-student approach requires high-quality teacher embeddings that are used as target.
\Cref{tab:teacher} shows the teacher model's performance on the VoxCeleb1-O trial set compared to the popular ECAPA-TDNN \cite{20_Desplanques_ecapa_tdnn}.
Here, it can be seen that the ResNet is able to achieve a comparable performance.
To keep a frame-wise resolution for loss computation, the ResNet architecture is the better choice for the student. Although any speaker embedding extractor can be used as teacher, we also chose a ResNet-based teacher to have matching configurations for teacher and student. 

\subsection{Necessity of teacher-student training}
First, we compare our proposed teacher-student training with a multi-speaker model that is trained from scratch using the same classification loss as the teacher, albeit with a \gls{PIT} objective. This second training approach is reminiscent of the way neural speech separation models are trained. Both models, the student of the T/S-approach and the extractor trained from scratch, use the same architecture described in \cref{sec:speaker_embeddings} and only differ in the loss used for training.
\begin{table}[bt]
    \caption{Multi-speaker verification performance for the student model and a multi-speaker embedding extractor trained with a classification loss on the VoxCeleb1-O m~vs.~m trials }
    \label{tab:mvm_loss}
    \centering
    \setlength{\tabcolsep}{4pt}

    \begin{tabular}{l c c c}
    \toprule
         Loss & Permutation & EER [\si{\percent}] & DCF\\
    \midrule
      $\mathcal{L}_{\mathrm{AAM}}$ & uPIT & 29.4 & 1.0\\ 
      $\mathcal{L}_{\mathrm{AAM}}$ & tPIT & 29.5 &  1.0\\ 
    \midrule
      $\mathcal{L}_{\mathrm{TS}}$ & uPIT & 19.9 & 0.88 \\ 
      $\mathcal{L}_{\mathrm{TS}}$ & tPIT &  \bfseries 14.1 &  \bfseries{0.74} \\ 
    \bottomrule
    \end{tabular}

\end{table}
\Cref{tab:mvm_loss} shows that training the speaker embedding extractor with a classification-based loss does not lead to a meaningful representation of the speaker embeddings independent of whether the permutation is solved per frame or per utterance. 
On the contrary, the teacher-student approach results in an \gls{EER} of \SI{14.1}{\percent}.
Using a training loss not on a frame-, but on an utterance-level did not work for either loss function.
This shows that the extraction of two embeddings from a speech mixture is considerably more intricate than the extraction of the speakers' speech signals from the mixture. This is probably because while the mixture is a linear superposition of the speech signals, it is not so for the embeddings.
Jointly learning a latent space and then projecting an utterance into this space leaves too many degrees of freedom for speech mixtures.
However, by providing this latent space through a pretrained speaker embedding extractor, the (student) model 
extracts significantly better speaker embeddings
that can be used to distinguish between the speakers. 
%

\begin{table}[bt]
\centering
\caption{Multi-speaker verification performance of the proposed model compared to a classical speaker embedding extractor on VoxCeleb mixtures. Model 1 and 2 indicate the model used on the first and second half of a trial pair.}
\setlength{\tabcolsep}{2pt}
\label{tab:voxceleb_results}
%
\robustify\bfseries
\sisetup{detect-weight=true,detect-inline-weight=text}
\begin{tabular}{l l c@{~~}S@{~~}S S@{}S }
\toprule
\multirow{2}{*}{Model 1} & \multirow{2}{*}{Model 2} &{\multirow{2}{*}{Scenario}}& \multicolumn{2}{c}{{any spk}} & \multicolumn{2}{c}{{per spk}} \\
\cmidrule(lr){4-5}\cmidrule(lr){6-7}
& & & {EER[\si{\percent}]} & {DCF} & {EER[\si{\percent}]} & {DCF}\\
\midrule
Teacher & Teacher & s~vs.~m & 18.2 & 0.57 & {-} & {-}\\
Teacher & Student & s~vs.~m & \bfseries 9.1 & \bfseries 0.46  & {-} & {-} \\
\midrule
Teacher & Teacher & m~vs.~m & 47.6 & 1.0 & {-} & {-}\\
Student & Student & m~vs.~m & \bfseries 15.3 & \bfseries 0.74 &\bfseries 14.1 &  \bfseries0.74\\
\bottomrule
\end{tabular}
\end{table}%

\subsection{Speaker embedding extraction for speech mixtures}
Next, we compare the advantage of using a distinct multi-speaker embedding
extractor over using only a single-speaker embedding extractor even in the presence of overlapping speech.
Here, both the teacher and proposed model are evaluated on the extended VoxCeleb1-O trial sets.
For the proposed model, the teacher is used to extract embeddings from single-speaker utterances, and the student for the embedding extraction from speech mixtures.
\Cref{tab:voxceleb_results} depicts the advantage of the combination of teacher and student over only using a single-speaker extractor. 
As expected, the proposed model outperforms the teacher in all scenarios.
For the \textit{s~vs.~m} trials, the teacher  is still able to re-identify speakers in a mixture to some degree, but its performance drops sharply when switching to the \textit{m~vs.~m} scenario where single-speaker regions are no longer available. On the contrary, the student is  able to achieve an \gls{EER} of \SI{15.3}{\percent}, which is, however, significantly higher compared to the \textit{s~vs.~m} scenario. In the \enquote{per spk} evaluation, the error rate decreases slightly, indicating that both extracted speaker embeddings can be used to identify the respective speakers.

To further quantify where the improvement of the student comes from and how disjoint the student embeddings are from each other, \cref{tab:crosstalker_leakage} evaluates how well the embeddings extracted from the mixture represent each active speaker.
This is done by calculating the cosine similarity between the speaker embeddings extracted from the mixture ($\hat{\mathbf{d}}_k$) and the respective teacher embeddings extracted from the clean, single-speaker signals ($\mathbf{d}_k$).  
This investigation shows that the teacher is able to accurately represent a single, the dominant, speaker in a mixture ($\mathbf{d}_y$), whereas the similarity to the other speaker stays low.
Compared to that, the student network shows a slightly lower, albeit still high average similarity to the dominant speaker, and an increased similarity to the second speaker.
Therefore, the student is able to extract both speakers, not only the dominant one as the teacher does.

\begin{table}[bt]
\centering
\caption{
Mean cosine similarity between different embeddings in a mixture and standard deviation over the trial examples (i.e. not model standard deviation). Similarities are ordered such that the higher score always is assigned to the first speaker.
}
\label{tab:crosstalker_leakage}
\setlength{\tabcolsep}{4pt}
\begin{tabular}{l c@{~~}c c@{}c@{~~}c}
\toprule 
& \multicolumn{2}{c}{Student (on $y$)} & \multicolumn{3}{c}{Teacher} \\
\cmidrule(lr){2-3}\cmidrule(lr){4-6}
Teacher & $\mathbf{\hat{d}}_1$ & $ \mathbf{\hat{d}}_2$ & $\mathbf{d}_1$&$\mathbf{d}_2$ &  $\mathbf{d}_y$ \\
\midrule
 $\mathbf{d}_1$ & .67 \footnotesize{$\pm$ .09} & .19 \footnotesize{$\pm$ .22} & 1 & -.09 \footnotesize{$\pm$ .12} & \bfseries .78 \footnotesize{$\pm$ .14} \\
 $\mathbf{d}_2$ & .13 \footnotesize{$\pm$ .19} & \bfseries .40 \footnotesize{$\pm$ .18} & -.09 \footnotesize{$\pm$ .12} & 1 & .22 \footnotesize{$\pm$ .18} \\
\bottomrule
\end{tabular}
\end{table}%

\subsection{Evaluation on AMI re-recordings}

Finally, the teacher-student model is evaluated on the AMI re-recordings from the SSLR dataset \cite{18_He_SSLR}.
Here, the trial sets for multi-speaker verification consist of all possible pairwise segment combinations per room. 
\Cref{tab:sslr_results} shows that our proposed single-channel system consistently outperforms the multi-channel baseline from \cite{21_He_multispeaker_embeddings}, which is a multi-channel system consisting of an MVDR beamformer and an x-vector embedding extractor, in the \textit{s~vs.~m} scenario, and in some cases also the hybrid model  proposed in  \cite{21_He_multispeaker_embeddings}. This effect becomes more pronounced for shorter segment lengths
and can also be seen in the \textit{m~vs.~m} scenario.
For longer segments, the direction of arrival information used in \cite{21_He_multispeaker_embeddings}  proves to be effective. Also noteworthy is that the SSLR dataset only contains \num{286} two-speaker segments in total, so more similar evaluation data may be necessary to solidify the results for the \textit{m~vs.~m} scenario.
Nevertheless, the results so far show that the proposed model achieves very good performance in the \textit{s~vs.~m} scenario on realistic monoaural speech data without any finetuning.

\begin{table}[bt]
\centering
\caption{EER (per spk) on the different SSLR trial sets for varying maximal segment lengths. * marks multi-channel models.}
\setlength{\tabcolsep}{3pt}
\sisetup{detect-weight=true,detect-inline-weight=text}
\label{tab:sslr_results}
\begin{tabular}{l c c c c c c}
\toprule
\multirow{2}{*}{Model} & \multicolumn{3}{c}{s~vs.~m} & \multicolumn{3}{c}{m~vs.~m}\\
\cmidrule(lr){2-4}\cmidrule(lr){5-7}

&  \SI{10}{\second} & \SI{5}{\second} & \SI{2}{\second} & \SI{10}{\second} & \SI{5}{\second} & \SI{2}{\second}\\
\midrule
MVDR+x-vector * \cite{21_He_multispeaker_embeddings}   & 10.8 & 12.7 & 20.4 & 12.2 & 14.3 & 21.7\\
Hybrid Model* \cite{21_He_multispeaker_embeddings}  & \bfseries ~~6.4 & 10.2 & 18.4 & \bfseries ~~6.3 & \bfseries 10.8  & 19.3\\
\midrule
Proposed & ~~8.4 & \bfseries ~~9.1 & \bfseries 14.2 & 10.9 & 11.7 & \bfseries 18.2\\
\bottomrule
\end{tabular}
\end{table}

\vspace{-1ex}
\section{Conclusions}
\label{sec:conclusion}
In this work, we proposed a system for speaker embeddings extraction from speech mixtures.
Using an embedding space defined by the teacher, a student embeddings extractor is learnt to cast a mixture input to embeddings in that space, representing the speakers present in the mixture.
Thus, speakers that have previously been active in some speech segments as the sole speaker, can be tested for their presence or absence in a mixture. The (re-)identification even works, however less reliably, if the speaker has never been active alone before. On re-recordings of the AMI dataset the proposed approach is able to outperform a multi-channel approach without additional finetuning, at least for short segments. 
As future work, we are planning to use the proposed multi-speaker embedding extractor to derive a speaker diarization system.
\vspace{-1ex}
\section{Acknowledgements}
Computational resources were provided by BMBF/NHR/PC2. Christoph Boeddeker was funded by DFG (German Research Foundation) under project no. 448568305.

\balance

\bibliographystyle{IEEEtran}
\bibliography{bibliography}

\end{document}